\documentclass[conference]{IEEEtran}
\IEEEoverridecommandlockouts

\newcommand{\mingjie}[1]{}

\def\namerm{Falcon}
\def\name{\textit{\namerm}\xspace}
\def\BibTeX{{\rm B\kern-.05em{\sc i\kern-.025em b}\kern-.08em
    T\kern-.1667em\lower.7ex\hbox{E}\kern-.125emX}}

\usepackage{amsmath}
\usepackage{amssymb}
\usepackage{xspace}
\usepackage{svg}
\usepackage{amsthm}

\usepackage{booktabs}
\usepackage{cite}
\usepackage[hidelinks]{hyperref}
\usepackage{multirow}
\usepackage{tabularx}
\usepackage{array}

\usepackage{threeparttable}
\usepackage{tablefootnote}

\usepackage{algorithm}
\usepackage{algorithmic}

\ifCLASSOPTIONcompsoc
  \usepackage[caption=false,font=normalsize,labelfont=sf,textfont=sf]{subfig}
\else
  \usepackage[caption=false,font=footnotesize]{subfig}
\fi

\usepackage{tikz}
\usetikzlibrary{decorations.pathreplacing,fit,intersections}

\newcommand{\figref}[1]{\figurename~\ref{#1}}

\theoremstyle{definition}

\begin{document}

\title{From Noisy Telemetry to Actionable Warnings: GPU Failure Prediction in Industrial Clusters}

\author{
\IEEEauthorblockN{
Yongqian Sun\IEEEauthorrefmark{2},
Run Zhu\IEEEauthorrefmark{2},
Wenwei Gu\IEEEauthorrefmark{2}\IEEEauthorrefmark{1},
Mengyao Li\IEEEauthorrefmark{2},
Shenglin Zhang\IEEEauthorrefmark{2},
Guanjin Wang\IEEEauthorrefmark{2}\\
Yang Zhang\IEEEauthorrefmark{3},
Xin Wu\IEEEauthorrefmark{3},
Linlin Han\IEEEauthorrefmark{3},
Feng Wang\IEEEauthorrefmark{3},
Xiaozhou Liu\IEEEauthorrefmark{3},
Yu Zhang\IEEEauthorrefmark{3}
}

\IEEEauthorblockA{
\IEEEauthorrefmark{2}\textit{Nankai University},
\{sunyongqian,wwgu,zhangsl\}@nankai.edu.cn\\
\{2211065,limengyao\}@mail.nankai.edu.cn,
fengwanhua.wgj@gmail.com
}

\IEEEauthorblockA{
\IEEEauthorrefmark{3}\textit{ByteDance},
\{zhangyang.329,wuxin.29,hanlinlin.intern,\\
wangfeng.ai,wangding.01,felix.zhang\}@bytedance.com
}

\thanks{\IEEEauthorrefmark{1}Wenwei Gu is the corresponding author.}
}

\maketitle

\setcounter{page}{1}
\thispagestyle{plain}
\pagestyle{plain}

\begin{abstract}
GPU clusters are critical infrastructure for AI services, but accurate and actionable GPU failure prediction remains a problem in production settings.
We study ticket-linked telemetry from a ByteDance GPU cluster and identify three obstacles: workload-confounded telemetry, heterogeneous fault precursors, and the gap between window-level predictions and actionable alerts. 
These findings motivate \name, a fault-specific warning framework combining missingness-aware temporal and peer-relative features, fault-specific learner selection, and an event policy based on thresholding, persistence, and cooldown.
On the test set, \name achieves the highest F1 among four baselines and reaches 70.6\% F1 on the best-performing fault type. 
Detected cases provide median lead times of 17.34--35.57 hours.
We further report a production deployment, where \name is calibrated toward high-precision alerts to reflect false-positive costs.
Together, these results show that fault-specific modeling improves early warning from noisy production GPU telemetry.
\end{abstract}


\begin{figure*}[!t]
  \centering
  \begin{minipage}[t]{0.49\textwidth}
    \centering
    \raisebox{0mm}{%
      \includegraphics[width=\linewidth]{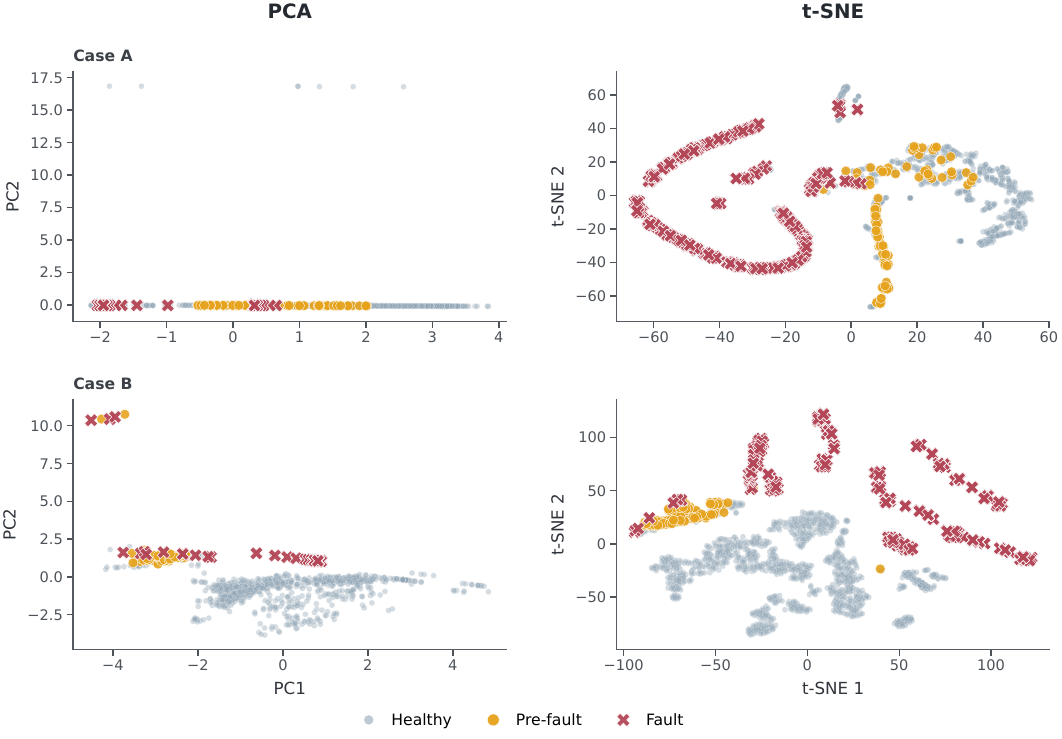}%
    }
    \vspace{-1mm}
    \centerline{\footnotesize (a) Projection evidence}
  \end{minipage}
  \hfill
  \begin{minipage}[t]{0.49\textwidth}
    \centering
    \raisebox{1.9mm}{%
      \includegraphics[width=\linewidth]{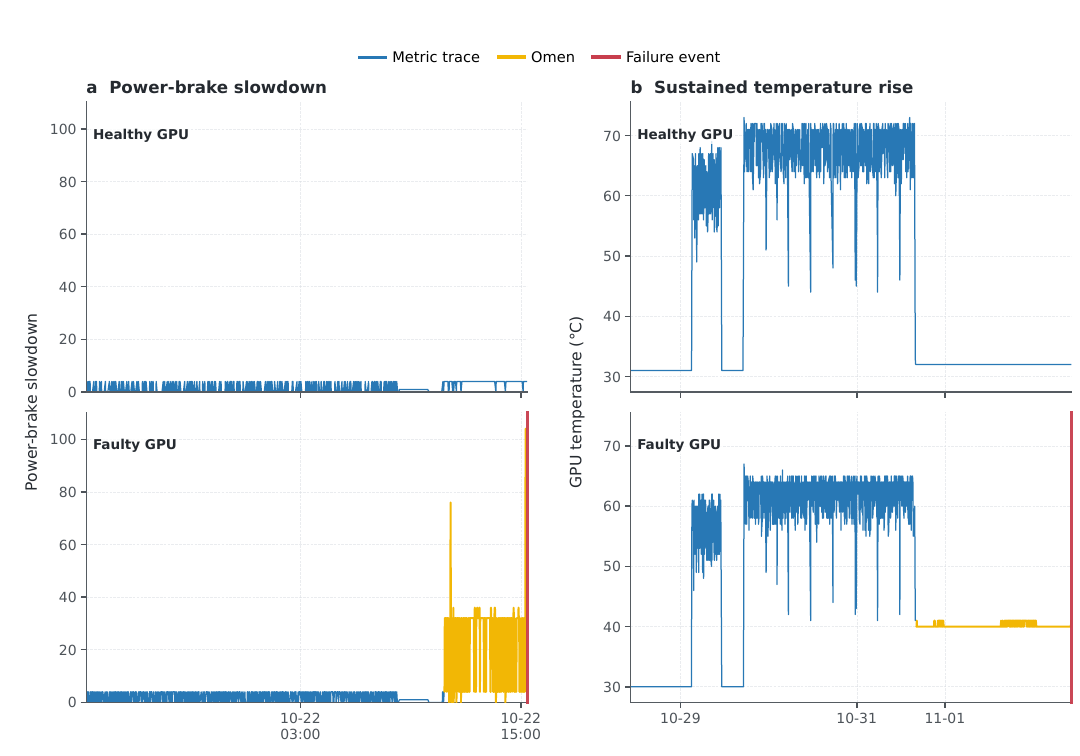}%
    }
    \vspace{-1mm}
    \centerline{\footnotesize (b) Trace evidence}
  \end{minipage}
  \caption{Empirical evidence for GPU fault predictability. PCA/t-SNE projections show that pre-fault states are not cleanly linearly separable from healthy states, while representative traces expose temporal shifts in power-brake slowdown and temperature.}
  \label{fig:predictability-evidence}
\end{figure*}
\section{Introduction}
\label{sec:introduction}

As modern software systems are increasingly driven by AI services, large language models, and agent workflows, their reliability increasingly depends on this AI infrastructure.
GPU faults therefore surface as software-level failures, including service interruptions, service-level objective (SLO) violations, failed training jobs, increased inference latency, and broken workflow execution.
At this scale, a device problem rarely stays local.
One unhealthy card can stall synchronization, trigger job recovery, occupy expensive capacity, or delay diagnosis across several infrastructure layers.
Large GPU clusters now carry workloads whose useful progress depends on thousands of tightly coupled accelerators, network links, power and cooling systems, drivers, and distributed runtimes~\cite{jiang2024megascale,meng2025astral,kokolis2025revisiting}.
GPU failure prediction is therefore a foundational problem for AI-native software reliability: risky devices must be identified early enough to inspect, drain, or replace them before the next ticket interrupts a workload.
Existing systems cover adjacent parts of this reliability problem.
Recent GPU degradation forecasting studies show that multivariate telemetry can contain pre-failure signals and use sliding-window health forecasting over GPU-cluster metrics~\cite{cai2025forecasting,liu2023predicting,nie2018machine}, but they primarily produce window-level health or failure scores rather than stable device-level warning events tied to fault families, tickets, and intervention horizons.
Diagnosis and mitigation systems such as Astral, Holmes, Minder, GREYHOUND, and Oobleck operate around runtime symptoms, faulty-machine localization, fail-slow mitigation, or resilient training~\cite{meng2025astral,yao2025holmes,deng2025minder,wu2025greyhound,jang2023oobleck}.
They motivate a complementary operations question: before a maintenance ticket, can ordinary monitoring telemetry become an early, precise, and actionable fault-specific warning?

The first obstacle is that production telemetry is not a direct fault signal.
Temperature, power, utilization, clocks, and NVLink or PCIe traffic are indirect health indicators and are strongly confounded by workload.
Optional counters may disappear for days, samples are not perfectly aligned, and healthy GPUs move among operating modes as jobs change.
The model must distinguish workload-driven healthy variation from pre-fault health changes under missing and misaligned telemetry.

Fault mechanisms are also heterogeneous.
Remapping failures may be preceded by remapping-state changes, ECC increments, or severe XID events; uncorrected-memory errors expose bursts of ECC or XID evidence; thermal faults show sustained temperature, throttling, or power-brake behavior; and power anomalies surface through abnormal power draw or braking states.
Because signal density, label volume, and lead time differ by fault, one architecture or global threshold is unlikely to serve all fault families.

Finally, prediction must become alerts.
Operators do not act on every window-level probability; they need stable warning events associated with a device, a fault family, and an intervention horizon.
Event formation is therefore part of the prediction problem: isolated spikes should be filtered, repeated warnings suppressed, and precision, recall, and lead time balanced under an operator-attention budget.
This also explains why the ByteDance remapping deployment uses a high-precision operating point rather than maximizing a window score alone.

We propose \name, short for Fault-specific Alerting for Large-scale GPU Clusters from Operations, as an analysis-driven warning framework for this setting.
\name addresses workload-confounded telemetry with missingness-aware temporal and peer-relative features, addresses fault heterogeneity by selecting learners separately for each fault, and addresses alert operationalization through thresholding, $N$-of-$K$ persistence, and cooldown.

This paper makes four contributions:
\begin{itemize}
  \item We formulate production GPU failure prediction as a fault-specific, event-level warning problem and analyze ticket-linked ByteDance telemetry to separate statistical separability from operational predictability.
  \item We present \name, which combines missingness-aware temporal features, peer-relative evidence, fault-specific learners, and event formation with thresholding, persistence, and cooldown.
  \item On four held-out fault types, \name achieves the best event-level F1 among the implemented baselines while providing 17.34--35.57 hours of median lead time.
  \item We report a production remapping-failure deployment at ByteDance with 0.800 converged-alert precision under a 2-day confirmation protocol.
\end{itemize}

\section{Empirical Study}
\label{sec:empirical}

We study anonymized telemetry and maintenance tickets from a production ByteDance GPU cluster covering more than 1,000 GPUs.
Each ticket records the affected GPU, fault type, occurrence time, and, when available, closing time.
The telemetry contains 67 raw fields, including 13 metadata fields and 54 monitoring fields such as temperature, power, clocks, utilization, PCIe/NVLink throughput, power-brake states, XID events, ECC counters, and row-remapping states.
The ticket corpus is broader than the prediction tasks: hardware faults, facility or environmental faults, and software-stack faults account for 41.05\%, 12.98\%, and 45.96\% of recorded tickets, respectively.

We do not treat every ticket type as a prediction target.
Some classes are not GPU-scoped, some have too few labels, and others lack continuous pre-fault telemetry.
We select four targets that are GPU-specific, have sufficient ticket-linked pre-fault telemetry, and expose plausible precursors for maintenance action: GPU Power Unknown Error, GPU Remapping Failure, GPU Temperature High, and GPU Uncorrected Error Over ByteDance Limit.
Other ticket types remain in the taxonomy but are excluded from model evaluation because they would require a different incident-classification setup.

We screen the selected targets for pre-fault evidence using KL divergence, two-sample KS tests, low-dimensional projections, and representative traces.
The checks give three findings.
First, some faults expose observable pre-fault signals.
At the 8-hour window, GPU memory temperature has mean and median KL values of 1.705 and 0.880, and GPU temperature has 1.785 and 0.813; neither metric has any job with KL below $10^{-1}$.
For remapping, power draw, memory temperature, GPU temperature, NVLink/PCIe throughput, and memory utilization also repeatedly pass the KS screen.
Second, workload variation prevents clean linear separability: PCA overlaps substantially, while t-SNE and traces reveal local shifts such as power-brake slowdown and sustained temperature rise (~\figref{fig:predictability-evidence}).
Third, precursor families differ by fault: temperature and power targets depend on dense thermal, power, and throttling channels, whereas remapping and uncorrected-memory targets rely more on sparse XID, ECC, and remapping-state evidence.

These findings motivate three design requirements.
The encoder must preserve temporal change and peer context rather than rely only on raw values.
Risk learning must be fault-specific instead of sharing one model and threshold across tickets.
Finally, separability does not guarantee deployability: window-level risk must be converted into stable alert events with thresholding, persistence, and cooldown before warning quality is evaluated.

\section{Methodology}
\label{sec:methodology}

\subsection{Overview and Problem Formulation}

\begin{figure}[t]
  \centering
  \includegraphics[width=0.49\textwidth]{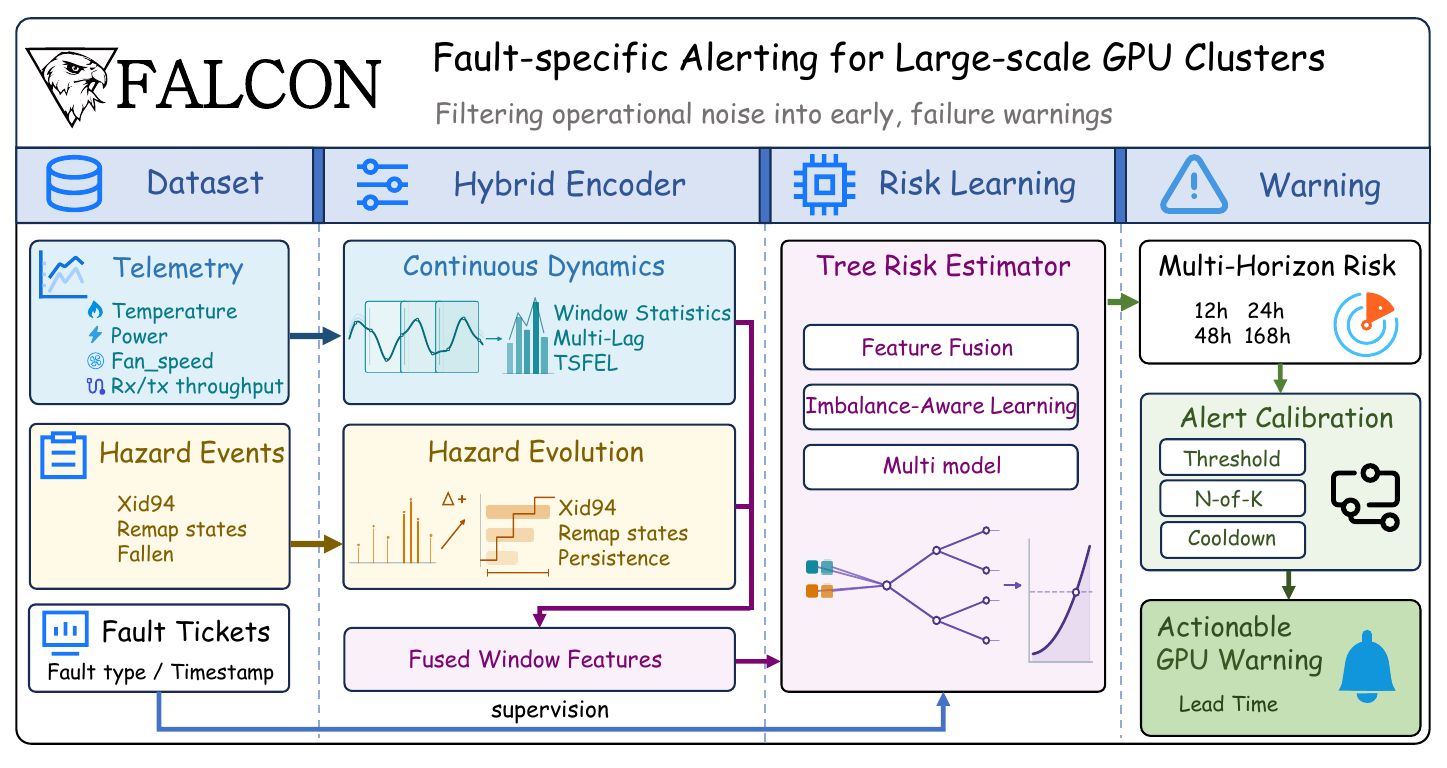}
  \caption{Fault-specific GPU warning framework. Dense telemetry and sparse hazard events are encoded into fused window features; a selected tree-based learner estimates risk; threshold, $N$-of-$K$ persistence, and cooldown convert window-level scores into actionable warning events.}
  \label{fig:framework}
\end{figure}

~\figref{fig:framework} summarizes \name. The framework maps directly to the three empirical challenges in Section~\ref{sec:empirical}. First, missingness-aware temporal and peer-relative encoding addresses indirect, incomplete, and workload-confounded telemetry. Second, fault-specific learner selection addresses heterogeneous precursor families and warning horizons. Third, alert-event formation addresses the operational gap between a window-level score and an operator-visible warning. \name therefore separates four decisions: define the target fault, encode recent telemetry and hazard evolution, learn a fault-specific risk score, and calibrate that score into a warning event.

For a fault type $f$, let $\mathbf{X}_{g,t-W:t}$ denote the telemetry of GPU $g$ in a window of length $W$ ending at time $t$. The model estimates
\begin{equation}
  p_f(g,t)=P\left(T_{g,f}\in(t,t+H]\mid \mathbf{X}_{g,t-W:t}\right),
  \label{eq:risk}
\end{equation}
where $T_{g,f}$ is the next ticket time for fault $f$ and $H$ is the prediction horizon. The score $p_f(g,t)$ is only a window-level risk estimate. A separate event policy decides whether successive scores become a valid warning within the intervention horizon.

\subsection{Temporal and Peer-Relative Feature Construction}

The feature encoder addresses incomplete and workload-confounded telemetry. Production GPU streams are sampled at one-minute intervals but remain partially missing and strongly affected by job phases. We align each GPU stream to the one-minute grid. Missing values in continuous telemetry are filled by linear interpolation before window aggregation. To make this imputation visible to the model, we also retain availability indicators, valid-sample counts, and missing ratios for each window. Temperature, power, clocks, utilization, and PCIe/NVLink throughput are then summarized by lagged statistics, changes, slopes, and persistence summaries. Count-like hazard fields, such as ECC and XID signals, are encoded through non-negative increments and event counts rather than smoothed averages.

Peer-relative features provide a local workload reference. For a target GPU, peers are the GPUs on the same IP that are running the same task at the same time; in our cluster, this group contains up to eight GPUs. When peer measurements are available, the encoder computes peer means, dispersions, z-scores, differences from peer means, ranks, and isolated-error indicators. These features ask whether one GPU is abnormal relative to its same-host, same-task context, which helps separate single-card degradation from collective workload shifts.
Table~\ref{tab:precursor-indicators} lists the ranked indicators used to construct temporal and peer-relative features.

\begin{table}[t]
\centering
\caption{Ranked precursor indicators used for temporal and peer-relative encoding.}
\label{tab:precursor-indicators}
\scriptsize
\setlength{\tabcolsep}{3pt}
\begin{tabularx}{\columnwidth}{@{}r>{\raggedright\arraybackslash}Xr>{\raggedright\arraybackslash}X@{}}
\toprule
Rank & Indicator & Rank & Indicator \\
\midrule
1 & Power draw & 6 & Memory utilization \\
2 & GPU memory temperature & 7 & PCIe RX throughput \\
3 & GPU temperature & 8 & PCIe TX throughput \\
4 & NVLink RX throughput & 9 & Power-brake slowdown \\
5 & NVLink TX throughput & 10 & SM clock \\
\bottomrule
\end{tabularx}
\end{table}

Sparse hazard streams are encoded as hardware-evolution signals. XID events, ECC increments, remapping-state changes, power-brake slowdown, event intervals, and consecutive hazardous states describe risk patterns that window averages can dilute. The deployed remapping configuration can additionally use TSFEL descriptors for dense signals~\cite{barandas2020tsfel}. Offline and deployed encoders are configuration-specific, but both produce fixed window-level representations for risk learning.

\subsection{Fault-Specific Risk Learning}

Fault-specific learning addresses heterogeneous mechanisms and warning horizons. We train one risk model per fault type because thermal faults, remapping failures, power anomalies, and uncorrected-memory errors expose different signals and tolerate different alert timing. Candidate learners include XGBoost~\cite{chen2016xgboost}, LightGBM~\cite{ke2017lightgbm}, CatBoost~\cite{prokhorenkova2018catboost}, Random Forest~\cite{breiman2001random}, and Isolation Forest~\cite{liu2008isolation}. These models fit heterogeneous tabular features containing imputed values, missingness indicators, sparse labels, and non-linear feature interactions.

For each target fault, validation selects the learner and event-policy parameters by event-level F1 over candidate learners, thresholds, $N$-of-$K$ rules, and cooldown values. The best configuration is frozen before held-out evaluation. Because positive ticket windows are rare, \name avoids accuracy and fixed 0.5 thresholds, using precision-recall-oriented selection for imbalanced prediction~\cite{saito2015precision}. The production remapping deployment uses a precision-oriented variant, as described in Section~\ref{sec:deploy}.

\subsection{Event-Level Warning Policy}

The event layer converts scores into maintenance-facing alerts. Given threshold $\tau_f$, a warning candidate is emitted only when at least $N_f$ of the latest $K_f$ scores exceed $\tau_f$; after emission, further warnings for the same GPU and fault are suppressed for cooldown $C_f$. Thresholding controls sensitivity, persistence filters isolated spikes, and cooldown prevents repeated alerts from one sustained condition. The policy is fault-specific because useful horizons and alert burden differ by fault.

Because this layer changes alert count and timing, it is validated with the learner rather than attached afterward. The pipeline loads the frozen learner, threshold, persistence rule, and cooldown to generate held-out warning events. A faulty case contributes one true positive if at least one warning occurs within the valid intervention horizon; alerts outside that horizon or on healthy cases are false positives. We therefore report case-level precision, recall, F1, and lead time after event formation, following prior time-series evaluation practice~\cite{tatbul2018precision}.

\section{Experiment}
\label{sec:experiment}

\subsection{Setup}

We evaluate \name on anonymized production telemetry and tickets from a ByteDance GPU cluster. The taxonomy uses the full corpus with more than 1,000 GPUs, while held-out experiments use a traceable subset of 313 GPUs sampled at one-minute intervals with telemetry and ticket artifacts. We study four ticket types: GPU Remapping Failure, GPU Power Unknown Error, GPU Temperature High, and GPU Uncorrected Error Over ByteDance Limit.

Data are split chronologically into training, validation, and test intervals with a 6:2:2 ratio. This matches deployment, where past behavior predicts future failures, and avoids leakage from non-temporal splits in which the same GPU or workload regime can expose future operating patterns during training.

The experiment uses two time scales. The window label uses a six-hour prediction horizon: a window is positive if the target ticket occurs within the next six hours, which gave the best validation performance among tested horizons. Window scores are then converted into events by threshold, $N$-of-$K$ persistence, and cooldown. Event-level evaluation uses a 48-hour intervention horizon selected by ByteDance operators for inspection, migration, isolation, or maintenance scheduling. Thus, six hours defines the learned risk target, whereas 48 hours defines when an alert remains actionable; it does not relax the prediction label. A faulty case is a true positive if any warning appears within 48 hours before its ticket. Missed faulty cases are false negatives; warnings on healthy cases are false positives. Lead time is measured from the first valid warning to the ticket and is reported only for detected cases. Offline experiments used Ubuntu/Python with standard gradient-boosting and scikit-learn libraries.

\subsection{Baselines}

We compare \name with four baselines. Liu et al.~\cite{liu2023predicting} represents GPU-specific failure prediction. Directly transferring this predictor to our setting is challenging because our features mainly describe short-term GPU telemetry dynamics, while the original study also uses broader contextual information such as GPU type, driver version, machine-room or rack location, and device aging. Without such context, the baseline has limited ability to explain abrupt or environment-related failures.

The LSTM baseline represents recurrent time-series anomaly detection~\cite{malhotra2015long}. LSTMs are suited to continuous temporal trends, but GPU fault precursors can be sparse, abrupt, or intermittent, such as sudden counter increments, transient ECC/remap-state changes, and short pre-fault windows with no clear trend. As a result, LSTM can cover some true failures and often achieves high recall, but it struggles to form precise decision boundaries under unstable short-window signals.

The Transformer baseline is trained from scratch on the same task~\cite{vaswani2017attention}; few and heterogeneous positive cases may limit the benefit of training sequence models from scratch. 

The ChatTime 7B Embedding baseline uses ChatTime 7B as a general embedding extractor with a lightweight classifier~\cite{DBLP:conf/aaai/Wang0W0ZWZL25}. It may capture generic numerical variation, while fault-specific structure may be needed to preserve physical semantics, cross-metric relations, and temporal context in GPU telemetry.

\subsection{RQ1: Effectiveness Across Fault Types}

Table~\ref{tab:rq1_baseline} reports the main event-level comparison. For compactness, Uncorrected Error Limit abbreviates GPU Uncorrected Error Over ByteDance Limit. \name achieves the highest F1 among the four baselines on all four faults. Temperature High is the most predictable target, reaching 0.632 precision, 0.800 recall, and 0.706 F1. Remapping Failure is also stable, with 0.611 precision and 0.500 F1. Power Unknown Error remains difficult because \name improves recall but still has low precision. Uncorrected Error reaches high precision.

\begin{table}[t]
\centering
\caption{Event-level comparison on the held-out test set.}
\label{tab:rq1_baseline}
\scriptsize
\setlength{\tabcolsep}{3.5pt}
\begin{tabular*}{\columnwidth}{@{\extracolsep{\fill}}llcccc@{}}
\toprule
Fault & Method & F1 & Precision & Recall & Median lead (h) \\
\midrule
\multirow{5}{*}{Remapping Failure}
& \textbf{\name} & \textbf{0.500} & \textbf{0.611} & 0.423 & 22.00 \\
& Transformer & 0.308 & 0.308 & 0.308 & 17.01 \\
& ChatTime & 0.229 & 0.182 & 0.308 & \textbf{27.50} \\
& Liu et al. & 0.071 & 0.067 & 0.077 & 1.06 \\
& LSTM & 0.495 & 0.343 & \textbf{0.885} & 2.75 \\
\midrule
\multirow{5}{*}{Power Unknown Error}
& \textbf{\name} & \textbf{0.333} & \textbf{0.219} & 0.700 & \textbf{35.57} \\
& Transformer & 0.267 & 0.200 & 0.400 & 23.77 \\
& ChatTime & 0.159 & 0.094 & 0.500 & 23.03 \\
& Liu et al. & 0.074 & 0.059 & 0.100 & 2.43 \\
& LSTM & 0.247 & 0.143 & \textbf{0.900} & 2.85 \\
\midrule
\multirow{5}{*}{Temperature High}
& \textbf{\name} & \textbf{0.706} & \textbf{0.632} & \textbf{0.800} & \textbf{23.02} \\
& Transformer & 0.323 & 0.312 & 0.333 & 22.03 \\
& ChatTime & 0.435 & 0.625 & 0.333 & 14.00 \\
& Liu et al. & 0.074 & 0.083 & 0.067 & 1.42 \\
& LSTM & 0.588 & 0.526 & 0.667 & 2.63 \\
\midrule
\multirow{5}{*}{Uncorrected Error Limit}
& \textbf{\name} & \textbf{0.444} & \textbf{0.667} & 0.333 & 17.34 \\
& Transformer & 0.111 & 0.067 & 0.333 & 14.50 \\
& ChatTime & 0.090 & 0.047 & \textbf{1.000} & \textbf{17.90} \\
& Liu et al. & 0.005 & 0.003 & 0.167 & 0.67 \\
& LSTM & 0.154 & 0.083 & \textbf{1.000} & 2.65 \\
\bottomrule
\end{tabular*}
\end{table}

The best \name learner also differs by fault: CatBoost is selected for Remapping Failure, Random Forest for Power Unknown Error, and LightGBM for Temperature High and Uncorrected Error. This pattern is consistent with the fault-specific design: the four ticket types expose different precursor patterns, sample sizes, and precision--recall trade-offs.

\subsection{RQ2: Ablation Study}

Table~\ref{tab:ablation} tests the two design choices most visible to operators: fault-specific learning and event-policy formation. ``All'' denotes one model for all faults. Removing $N$-of-$K$ persistence often increases sensitivity but also raises false positives. For Remapping Failure, recall rises from 0.423 to 0.692, but F1 drops from 0.500 to 0.409. Removing cooldown degrades all four faults, confirming that alert deduplication is part of the predictive system rather than only a post-processing convenience. A shared model often increases recall, but its precision drops, showing that one learner does not fit all fault mechanisms.

\begin{table}[t]
\centering
\caption{Ablation results on the held-out test set.}
\label{tab:ablation}
\scriptsize
\setlength{\tabcolsep}{3.5pt}
\begin{tabular*}{\columnwidth}{@{\extracolsep{\fill}}llcccc@{}}
\toprule
Fault & Variant & F1 & Precision & Recall & Median lead (h) \\
\midrule
\multirow{4}{*}{Remapping Failure}
& Full & \textbf{0.500} & \textbf{0.611} & 0.423 & 22.00 \\
& w/o persistence & 0.409 & 0.290 & 0.692 & 21.02 \\
& w/o cooldown & 0.278 & 0.500 & 0.192 & 20.03 \\
& All & 0.339 & 0.214 & \textbf{0.808} & \textbf{25.00} \\
\midrule
\multirow{4}{*}{Power Unknown Error}
& Full & \textbf{0.333} & 0.219 & \textbf{0.700} & \textbf{35.57} \\
& w/o persistence & 0.267 & 0.200 & 0.400 & 11.89 \\
& w/o cooldown & 0.182 & \textbf{1.000} & 0.100 & 4.93 \\
& All & \textbf{0.333} & 0.219 & \textbf{0.700} & \textbf{35.57} \\
\midrule
\multirow{4}{*}{Temperature High}
& Full & \textbf{0.706} & \textbf{0.632} & 0.800 & 23.02 \\
& w/o persistence & 0.552 & 0.571 & 0.533 & 9.02 \\
& w/o cooldown & 0.167 & 0.143 & 0.200 & 9.70 \\
& All & 0.500 & 0.351 & \textbf{0.867} & \textbf{24.38} \\
\midrule
\multirow{4}{*}{Uncorrected Error Limit}
& Full & \textbf{0.444} & \textbf{0.667} & 0.333 & \textbf{17.34} \\
& w/o persistence & 0.250 & 0.200 & 0.333 & 14.30 \\
& w/o cooldown & 0.286 & 0.250 & 0.333 & 5.00 \\
& All & 0.273 & 0.187 & \textbf{0.500} & 12.13 \\
\bottomrule
\end{tabular*}
\end{table}

\subsection{RQ3: Lead-Time Analysis}

\name provides 17.34--35.57 hours of median lead time among detected cases, within the 48-hour actionable horizon for inspection, isolation, or maintenance scheduling. Lead time must be interpreted with alert quality. ChatTime gives longer median lead time for Remapping Failure but only 0.182 precision; on Uncorrected Error it detects all faulty cases but raises 121 false-positive cases, reducing precision to 0.047. Earlier warning is useful only under a comparable precision or alert budget. Overall, \name improves the precision--recall trade-off while preserving actionable lead time.

\section{Deployment}
\label{sec:deploy}

In production, the system uses a precision-oriented strategy tailored to operational constraints. Unlike the offline study, deployment must account for per-alert costs: workload migration, scheduling changes, diagnostics, stress validation, and manual inspection. The objective therefore shifts toward high alert precision because each false positive triggers costly operational work. Accordingly, deployment calibration uses a precision-oriented objective rather than the offline F1 objective, and the model output is treated as an operational trigger rather than a purely statistical score.

The deployed pipeline couples periodic retraining with online inference. It aggregates temperatures, power draw, throttling indicators, XID counts, uncorrectable ECC errors, and remapped-row metrics, then derives TSFEL-based~\cite{barandas2020tsfel} temporal descriptors and rule-based failure features over sliding windows. Retraining uses a rolling 100-day window, with 80 days for training and 20 days for validation. The threshold is selected on the precision-recall curve to maximize precision under a minimum-recall constraint, and a model is promoted only if validation precision satisfies the deployment criterion. At inference time, the system uses the selected feature subset and persisted threshold to preserve training-serving consistency.

Online inference runs every 20 minutes. Each run retrieves the latest model artifacts and applies the same aggregation, imputation, sliding-window feature construction, and probabilistic scoring to new telemetry. To suppress repeated alerts during sustained degradation, convergence keeps only the earliest high-confidence alert for each IP within 48 hours. This converts continuous risk scores into a tractable alert stream aligned with ticketing, migration, and diagnostic workflows.

\begin{figure}
  \begin{center} 
  \includegraphics[width=\columnwidth]{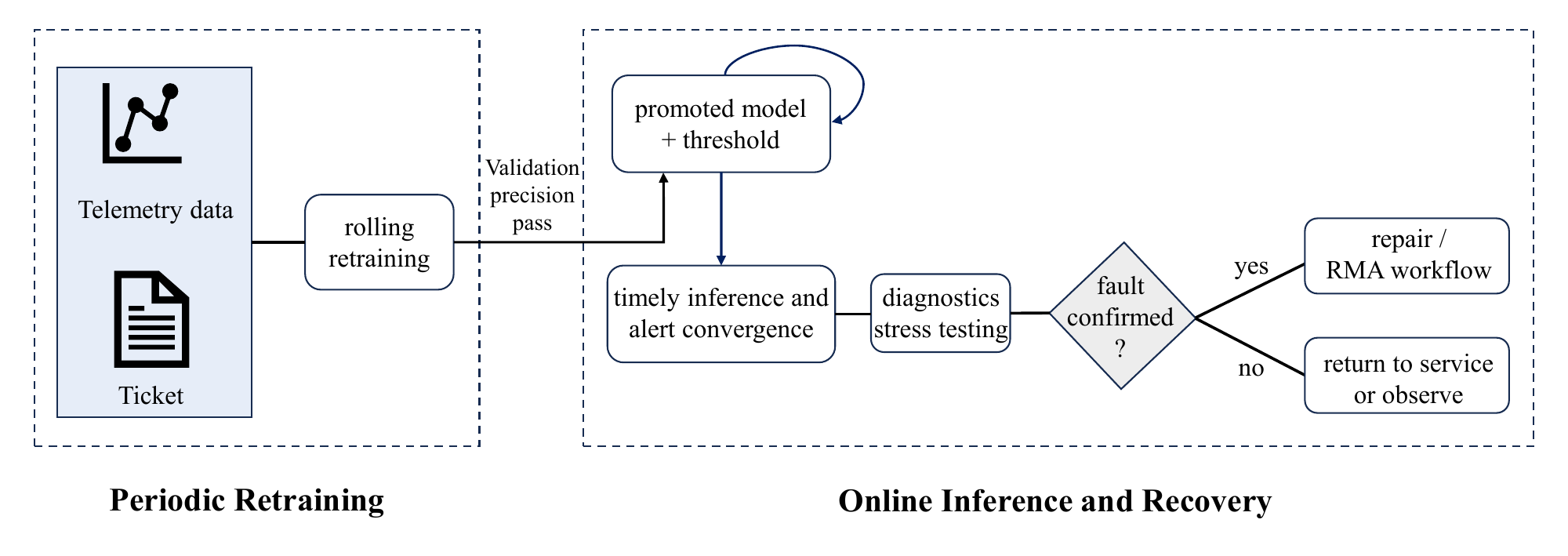}
  \caption{\name production deployment workflow.}\label{fig:deployment}
  \end{center}
\end{figure}

A high-confidence alert is not treated as definitive evidence of hardware failure. It triggers proactive intervention: workloads are migrated away from the suspected host, and the alert is checked through diagnostics, stress testing, and hardware-state inspection. Confirmed faults enter repair or return merchandise authorization (RMA); unconfirmed hosts return to service or remain under observation. This couples model-based prediction, risk isolation, and human-in-the-loop validation.

The deployment ran for one month on a production GPU cluster covering more than 10,000 GPUs. After convergence, the system generated 50 alerts, of which 40 were confirmed by engineers. 
Because the complete denominator of production remapping failures is confidential, we report precision but not recall; the result therefore characterizes alert confidence rather than overall failure coverage.
These results show that the precision-oriented strategy can produce a tractable high-confidence alert stream and connect offline prediction to practical intervention through migration, diagnosis, and RMA decisions.

\section{Related Work}
\label{sec:related}

\subsection{GPU Cluster Reliability and Diagnosis}

GPU-cluster studies address reliability, diagnosis, and mitigation. Astral, large-scale cluster studies, and prior HPC work characterize infrastructure behavior and failures~\cite{meng2025astral,kokolis2025revisiting}. 
Runtime and incident-management systems localize training irregularities, detect fail-slow behavior, identify faulty machines, improve training resilience, or automate multimodal anomaly detection, failure triage, and root-cause localization around observed runtime symptoms~\cite{yao2025holmes,wu2025greyhound,deng2025minder,jang2023oobleck,Sun2025TrioXpertAA}. 
\name addresses a different operating point: infrastructure-level telemetry and case-level warning before maintenance tickets occur.

\subsection{GPU Failure Prediction and Predictive Maintenance}

\emph{Forecasting Machine Degradation of GPU Clusters} is the most closely related GPU-specific predictive-maintenance study~\cite{cai2025forecasting}. 
It uses multivariate telemetry to forecast GPU-cluster degradation, showing that production telemetry can contain pre-failure information. 
\emph{Predicting GPU Failures With High Precision Under Deep Learning Workloads} studies production GPU failure prediction and uses ensemble and sliding-training techniques to improve precision and stability~\cite{liu2023predicting}. 
Machine-learning models for GPU error prediction in large-scale HPC systems provide another related predictive setting~\cite{nie2018machine}.
%Outside GPU clusters, Xu et al. predict disk errors in cloud systems to improve service availability~\cite{xu2018disk}; Falcon shares this operational motivation but targets heterogeneous GPU fault families and converts scores into fault-specific, case-level warning events.
%We further compare with LSTM anomaly detection~\cite{malhotra2015long}, Transformer sequence modeling~\cite{vaswani2017attention}, and ChatTime time-series representations~\cite{DBLP:conf/aaai/Wang0W0ZWZL25}. 
\name differs by defining fault-specific ticket targets and evaluating GPU-level warning events with persistence, cooldown, precision, recall, F1, and lead time.

\section{Conclusion}
\label{sec:conclusion}
Actionable GPU warning requires more than high risk scores: production telemetry is incomplete, workload-entangled, fault-specific, and costly to operationalize. Falcon combines fault qualification, temporal and peer-relative evidence, fault-specific learners, and a validated threshold–persistence–cooldown policy. Across four held-out faults, it improves event-level F1 over the implemented baselines, while the one-month deployment produces a high-confidence alert stream for a cluster with more than 10,000 GPUs.

\section{Acknowledgement}

This work is supported by the National Natural Science Foundation of China (62272249, 62302244), the Fundamental Research Funds for the Central Universities (XXX-63253249), and the Tianjin Key Research and Development Program (Grant No. 25YFYFFG00690).

\bibliographystyle{IEEEtran}
\bibliography{bibliography}

@inproceedings{jiang2024megascale,
  title={MegaScale: Scaling large language model training to more than 10,000 GPUs},
  author={Jiang, Ziheng and Lin, Haibin and Zhong, Yinmin and Huang, Qi and Chen, Yangrui and Zhang, Zhi and Peng, Yanghua and Li, Xiang and Xie, Cong and Nong, Shibiao and others},
  booktitle={21st USENIX Symposium on Networked Systems Design and Implementation (NSDI 24)},
  pages={745--760},
  year={2024}
}

@inproceedings{meng2025astral,
  title={Astral: A datacenter infrastructure for large language model training at scale},
  author={Meng, Qingkai and Zheng, Hao and Zhang, Zhenhui and Lao, ChonLam and Huang, Chengyuan and Li, Baojia and Zhu, Ziyuan and Lu, Hao and Dang, Weizhen and Lin, Zitong and others},
  booktitle={Proceedings of the ACM SIGCOMM 2025 Conference},
  pages={609--625},
  year={2025}
}

@inproceedings{kokolis2025revisiting,
  title={Revisiting reliability in large-scale machine learning research clusters},
  author={Kokolis, Apostolos and Kuchnik, Michael and Hoffman, John and Kumar, Adithya and Malani, Parth and Ma, Faye and DeVito, Zachary and Sengupta, Shubho and Saladi, Kalyan and Wu, Carole-Jean},
  booktitle={2025 IEEE International Symposium on High Performance Computer Architecture (HPCA)},
  pages={1259--1274},
  year={2025},
  organization={IEEE}
}

@inproceedings{cai2025forecasting,
  title={Forecasting machine degradation of GPU Clusters},
  author={Cai, Shengnan and Nie, Shuxin and Chen, Zhehui and Gulalkari, Nupur and Vanica, George and Jain, Chetna and Sankaran, Sethuraman},
  booktitle={Machine Learning for Systems 2025},
  year={2025}
}

@inproceedings{liu2023predicting,
  title={Predicting gpu failures with high precision under deep learning workloads},
  author={Liu, Heting and Li, Zhichao and Tan, Cheng and Yang, Rongqiu and Cao, Guohong and Liu, Zherui and Guo, Chuanxiong},
  booktitle={Proceedings of the 16th ACM International Conference on Systems and Storage},
  pages={124--135},
  year={2023}
}

@inproceedings{nie2018machine,
  title={Machine learning models for GPU error prediction in a large scale HPC system},
  author={Nie, Bin and Xue, Ji and Gupta, Saurabh and Patel, Tirthak and Engelmann, Christian and Smirni, Evgenia and Tiwari, Devesh},
  booktitle={2018 48th Annual IEEE/IFIP International Conference on Dependable Systems and Networks (DSN)},
  pages={95--106},
  year={2018},
  organization={IEEE}
}

@inproceedings{yao2025holmes,
  title={Holmes: Localizing irregularities in $\{$LLM$\}$ training with mega-scale $\{$GPU$\}$ clusters},
  author={Yao, Zhiyi and Hu, Pengbo and Miao, Congcong and Jia, Xuya and Liang, Zuning and Xu, Yuedong and He, Chunzhi and Lu, Hao and Chen, Mingzhuo and Li, Xiang and others},
  booktitle={22nd USENIX Symposium on Networked Systems Design and Implementation (NSDI 25)},
  pages={523--540},
  year={2025}
}

@inproceedings{deng2025minder,
  title={Minder: Faulty machine detection for large-scale distributed model training},
  author={Deng, Yangtao and Shi, Xiang and Jiang, Zhuo and Zhang, Xingjian and Zhang, Lei and Zhang, Zhang and Li, Bo and Song, Zuquan and Zhu, Hang and Liu, Gaohong and others},
  booktitle={22nd USENIX Symposium on Networked Systems Design and Implementation (NSDI 25)},
  pages={505--521},
  year={2025}
}

@inproceedings{wu2025greyhound,
  title={$\{$GREYHOUND$\}$: Hunting $\{$Fail-Slows$\}$ in $\{$Hybrid-Parallel$\}$ Training at Scale},
  author={Wu, Tianyuan and Wang, Wei and Yu, Yinghao and Yang, Siran and Wu, Wenchao and Duan, Qinkai and Yang, Guodong and Wang, Jiamang and Qu, Lin and Zhang, Liping},
  booktitle={2025 USENIX Annual Technical Conference (USENIX ATC 25)},
  pages={731--747},
  year={2025}
}

@inproceedings{jang2023oobleck,
  title={Oobleck: Resilient distributed training of large models using pipeline templates},
  author={Jang, Insu and Yang, Zhenning and Zhang, Zhen and Jin, Xin and Chowdhury, Mosharaf},
  booktitle={Proceedings of the 29th Symposium on Operating Systems Principles},
  pages={382--395},
  year={2023}
}

@article{barandas2020tsfel,
  title={TSFEL: Time series feature extraction library},
  author={Barandas, Mar{\'\i}lia and Folgado, Duarte and Fernandes, Let{\'\i}cia and Santos, Sara and Abreu, Mariana and Bota, Patr{\'\i}cia and Liu, Hui and Schultz, Tanja and Gamboa, Hugo},
  journal={SoftwareX},
  volume={11},
  pages={100456},
  year={2020},
  publisher={Elsevier}
}

@inproceedings{chen2016xgboost,
  title={Xgboost: A scalable tree boosting system},
  author={Chen, Tianqi and Guestrin, Carlos},
  booktitle={Proceedings of the 22nd acm sigkdd international conference on knowledge discovery and data mining},
  pages={785--794},
  year={2016}
}

@article{ke2017lightgbm,
  title={Lightgbm: A highly efficient gradient boosting decision tree},
  author={Ke, Guolin and Meng, Qi and Finley, Thomas and Wang, Taifeng and Chen, Wei and Ma, Weidong and Ye, Qiwei and Liu, Tie-Yan},
  journal={Advances in neural information processing systems},
  volume={30},
  year={2017}
}

@article{prokhorenkova2018catboost,
  title={CatBoost: unbiased boosting with categorical features},
  author={Prokhorenkova, Liudmila and Gusev, Gleb and Vorobev, Aleksandr and Dorogush, Anna Veronika and Gulin, Andrey},
  journal={Advances in neural information processing systems},
  volume={31},
  year={2018}
}

@article{breiman2001random,
  title={Random forests},
  author={Breiman, Leo},
  journal={Machine learning},
  volume={45},
  number={1},
  pages={5--32},
  year={2001},
  publisher={Springer}
}

@inproceedings{liu2008isolation,
  title={Isolation forest},
  author={Liu, Fei Tony and Ting, Kai Ming and Zhou, Zhi-Hua},
  booktitle={2008 eighth ieee international conference on data mining},
  pages={413--422},
  year={2008},
  organization={IEEE}
}

@article{saito2015precision,
  title={The precision-recall plot is more informative than the ROC plot when evaluating binary classifiers on imbalanced datasets},
  author={Saito, Takaya and Rehmsmeier, Marc},
  journal={PloS one},
  volume={10},
  number={3},
  pages={e0118432},
  year={2015},
  publisher={Public Library of Science}
}

@article{tatbul2018precision,
  title={Precision and recall for time series},
  author={Tatbul, Nesime and Lee, Tae Jun and Zdonik, Stan and Alam, Mejbah and Gottschlich, Justin},
  journal={Advances in neural information processing systems},
  volume={31},
  year={2018}
}

@inproceedings{malhotra2015long,
  title={Long short term memory networks for anomaly detection in time series},
  author={Malhotra, Pankaj and Vig, Lovekesh and Shroff, Gautam and Agarwal, Puneet and others},
  booktitle={Proceedings},
  volume={89},
  number={9},
  pages={94},
  year={2015}
}

@article{vaswani2017attention,
  title={Attention is all you need},
  author={Vaswani, Ashish and Shazeer, Noam and Parmar, Niki and Uszkoreit, Jakob and Jones, Llion and Gomez, Aidan N and Kaiser, {\L}ukasz and Polosukhin, Illia},
  journal={Advances in neural information processing systems},
  volume={30},
  year={2017}
}

@inproceedings{DBLP:conf/aaai/Wang0W0ZWZL25,
  title={Chattime: A unified multimodal time series foundation model bridging numerical and textual data},
  author={Wang, Chengsen and Qi, Qi and Wang, Jingyu and Sun, Haifeng and Zhuang, Zirui and Wu, Jinming and Zhang, Lei and Liao, Jianxin},
  booktitle={Proceedings of the AAAI Conference on Artificial Intelligence},
  volume={39},
  number={12},
  pages={12694--12702},
  year={2025}
}

@article{Sun2025TrioXpertAA,
  title={TrioXpert: An Automated Incident Management Framework for Microservice System},
  author={Sun, Yongqian and Luo,Yu and Wen, Xidao and Yuan, Yuan and Nie, Xiaohui and Zhang, Shenglin and Liu, Tong and Luo, Xi},
  journal={2025 40th IEEE/ACM International Conference on Automated Software Engineering (ASE)},
  year={2025},
  pages={3239-3250},
}

\end{document}